# A new topological insulator - β-InTe strained in the layer plane

F.M. Hashimzade[a], D.A. Huseinova[a], Z.A. Jahangirli[a,b], N.T. Mamedov[a],

B.H. Mehdiyev[a]

[a]Institute of Physics, National Academy of Sciences of Azerbaijan, AZ 1143, Baku, Azerbaijan

[b]Azerbaijan Technical University, pr. H. Javid 25, AZ1073, Baku, Azerbaijan

**ABSTRACT**

We have investigated the band structure of the bulk crystal and the (001) surface of the β-InTe layered crystal subjected to biaxial stretching in the layer plane. The calculation has been carried out using the full-potential linearized augmented plane wave method (FP LAPW) implemented in WIEN2k. It has been shown that at the strain $\Delta a/a = 0.06$, where a is the lattice parameter in the layer plane, the band gap in the electronic spectrum collapses. With further strain increase a band inversion occurs. The inclusion of the spin-orbit interaction reopens the gap in the electronic spectrum of a bulk crystal, and our calculations show that the spectrum of the surface states has the form of a Dirac cone, typical for topological insulators.

1.  **Introduction**

A search for and a study of the new materials with special properties, enabling their use in quantum computers and spintronics, is currently one of the main directions in physics of semiconductors. Among the new promising materials is the recently discovered class of topological insulators (TI), which are insulators in the bulk and possess protected by time reversal symmetry metallic conductivity on the surface, due to a strong spin-orbit interaction [1,2]. These surface states form a Dirac cone in the time-reversal invariant momenta (TRIM) points of the brillouin zone (BZ), where the direction of the spin is locked to the momentum, so that the states with opposite spins propagate in opposite directions. The time reversal symmetry leads to the absence of the backscattering of electrons on defects. Therefore, due to the gapless states on the surface or at the boundary with an ordinary insulator the flow of spin-polarized current is possible practically without energy loss. The compounds with TI properties must be insulators in the bulk, have strong spin-orbit coupling with energy commensurable with the energy gap, and have inverted band gap.

Recently, Y.Ma et al. [3] predicted theoretically that β-InSe can transit into the topological insulator state under a two-axis strain in the (001) plane. Unlike the common understanding of the mechanism of band inversion in three-dimensional TI, here the inversion is not caused by the spin-orbit coupling, but, instead, is due to the lattice deformation caused by an external tension. In this case the bands with even and odd symmetry around band gap swap places after the collapse of the energy gap and a further increase of strain. Spin-orbit coupling is still necessary for reopening of the energy gap, which leads to the TI state. According to these authors' results, at a 6% strain β-InSe turns into a TI with the energy gap of 121 meV. However, it is known that the band gap in β-InSe is about 1.45 eV [4], and the method of local density approximation (LDA) used by the authors of [3] gives a strongly under-estimated value of the energy gap. This led to an overly optimistic prediction degree of the deformation necessary for achieving the TI state. Realistically, as we show below, achieving the effect described by the authors requires much higher strain.

Nevertheless, the work in [3] has stimulated our search for other suitable objects for TI engineering via strain. In [5], using the first-principles calculations, it has been shown that the total energy of InTe in the hexagonal and tetragonal phases is almost the same at zero temperature and pressure, and, therefore, this crystal can exist in a metastable hexagonal phase of β-InTe.

The crystals of β-InTe and β-InSe are isostructural and have a layered structure of crystal lattice. The layers consist of a package of 4 alternating atomic planes, Te-In-Te-In. In the neighboring layers the atoms of In are above the atoms of Te. The space group symmetry is P63/mmc (No. 194).

In the current work the electronic spectrum of the bulk crystal and of the (001) surface of β-InTe has been investigated from the first principles. It has been shown that this crystal transits into a TI under a two-axis strain in the layer plane. In the first-principles calculations, in order to obtain the correct gap, we have used a modified Becke-Johnson (mBJ) potential. Our calculations show that the gap collapses at a 6% strain, and at further strain increase a band inversion (BI) occurs. For gap reopening one has to take into account the spin-orbit coupling, which, at a 9% strain, gives an energy gap of the order of 169 meV.

## 2. **Method of calculation.**

In our calculations we use the FP LAPW method implemented in WIEN2k package [6]. Preliminarily, the equilibrium values of the lattice parameters and the positions of atoms in the unit cell were calculated. For this, we used the ABINIT package [7]. The equilibrium values of the lattice parameters were determined by minimizing the total energy with an accuracy of $10^{-6}$ eV per unit cell. The equilibrium atomic positions in a unit cell were determined by the minimization of Hellmann–Feynman forces using the experimental structure as a starting point. The process of minimization continued until the force moduli became less than $10^9$ eV/m. The total energy was calculated using a 12×12×3 Monkhorst-Pack k-point grid [8]. We used the norm-conserving Hartwigsen-Goedecker-Hutter pseudo-potential [9] and the plane wave basis with the kinetic energy up to 1500 eV.

In the calculations of the electronic spectrum of the bulk and surface state using WIEN2k package the exchange-correlation potential within the general gradient approximation (GGA) was calculated using scheme suggested by Perdew et al. [10]. Surface electronic structure calculations are based on supercell using bulk relaxed parameters containing 8 layers with vacuum of 68.9 Å. After optimization structural parameters of beta-InTe are: $a = 4.203$ Å, $c = 17.235$ Å, $z_{In} = 0.1728$, $z_{Te} = 0.09441$. In our calculations, the convergence parameter RmtKmax was set to 7.0. The value of the parameter Kmax determines the size of the basis sets in the calculations. Inside atomic spheres the partial waves were expanded up to $lmax = 10$. Integrations in reciprocal space were performed using the tetrahedron method with 1000 k points in the irreducible part of the first Brillouin zone. The muffin-tin sphere radii $Rmt = 2.5$ a.u. were used for In and Te. The cut-off energy for separation of the core and valence states was set to -6.0 Ry. Spin-orbit interactions have been considered via a second variational step using the scalar-relativistic eigenfunctions as basis.

3. **Results and discussion.**

We have calculated the (bulk) band structure of β-InTe under the strain $\varepsilon = \Delta a/a$ in the interval from 0 to 0.11, where $a$ is the lattice parameter in the layer plane. The calculated (bulk) band structure of β-InTe at $\varepsilon = 0.05$ without and with spin-orbit coupling (SOC) is shown in Figs. 1(a) and 1(b).

As one can see from Fig. 1, strained β-InTe with $\varepsilon = 0.05$, without SOC, is a direct band gap semiconductor with the energy gap of 0.314 eV located at the centre of the Brillouine zone. With SOC the energy gap become 0.225 eV. By including SOC, band inversion of the conduction band minimum (CBM) and valence band maximum (VBM) does not occur. So, the spin-orbit interaction in b-InTe is not strong enough to induce the band inversion. According to [5], CBM and VBM are formed by s and p states of In and Te atoms. Since under strain in the layer plane the crystal simultaneously shrinks in the direction perpendicular to the layers, we have also show the dependence of parameter c upon ε in Fig. 3. One can see from Fig. 3 that the character of this dependence changes sharply near ε=0.06. We will see below that this correlates with the transition of the semiconductor to a semi-metal.

Fig. 4 shows the dependence of the energy gap at Γ point upon the degree of the strain. As one can see from the figure, the energy gap continuously narrows as the degree of strain increases. At the critical value of $\varepsilon = 0.06$, the band gap closes. As the strain increases further, the conduction band and the valence band cross, and the energy gap becomes negative. Because these bands, according to [5], have opposite parity, one can say that at the strain with $\varepsilon > 0.06$, a band inversion takes place. Figs. 5(a) and 5(b) show the band structure of β-InTe without and with SOC for ε=0.09. As one can see from Fig. 5(a), without SOC, the bands cross along the Γ-K and Γ–M lines, and β-InTe is a semi-metal. With SOC, there is a gap of 169 eV, which exceeds the energy scale at the room temperature. The gap opens because of the symmetry reduction and in accordance with the non-crossing rule anti-crossing of terms with the same symmetry. According to [3], the change in the parity of valence band states at Γ point leads to a transition into TI phase with the topological invariant Z2 = 1. To verify the topological nature of strained β-InTe with $\varepsilon > 0.06$ we have calculated the surface band structure of β-InTe at $\varepsilon = 0.07$ using the slab method containing super-cells repeated periodically in the direction perpendicular to the (001) plane.

Fig. 7 shows the surface band structure, together with the projection of the bulk band structure onto the surface Brillouine zone depicted in Fig. 6. As one can see from Fig. 7, the surface state spectrum has the shape of a Dirac cone located within the gap of the bulk state. This proves the transition of β-InTe semiconductor crystal subjected to strain with $\varepsilon > 0.06$ into a TI.

4. **Conclusion**

The band structures of bulk and surface states of β-InTe hexagonal crystal have been calculated. It has been shown that a two-axis strain of the layered crystal in the layer plane leads to a transition of this semiconductor into a TI state. The transition occurs as a result of band inversion after the energy band collapses at the strain above the critical value $\varepsilon > 0.06$.

Taking the SOC into account leads to a gap opening of about 169 meV in the bulk state spectrum, whereas the surface state spectrum remains gapless. Thus, strained β-InTe can be a suitable material for spintronics at room temperature.

**Acknowledgments**

The authors would like to thank Alexey Bondyakov (Joint Institute for Nuclear Research, Dubna, Russia), as well as the entire staff of the AZGRID Data Center of the Institute of Physics of the ANAS for providing resources and technical support for theoretical calculations. This work was supported by the Science Development Foundation under Grant EIF-KETPL-2-2015-1(25)-56/02/1.

**Appendix**

We have mentioned above that, according to [5], β-InTe is meta-stable at zero temperature and pressure. An interesting question is whether the hexagonal structure of this crystal is more stable than the tetragonal one. To answer this question we have calculated the equations of state for InTe in both crystalline structures. The parameters of tetragonal crystalline structure were taken from [11] and optimised using ABINIT package [7]. The computed total energies as functions of the unit cell volume for hexagonal and tetragonal InTe crystals are shown in Figs. 8 and 9. The solid lines in these figures were obtained by fitting to Birch-Murnaghan equation of state [12,13]. Using the Birch-Murnaghan equation, we have compared the enthalpies of the hexagonal and tetragonal InTe crystals over the range of strain degrees that induce a phase transition of β-InTe into a TI. Over the entire range of strain with $\varepsilon > 0.06$ the hexagonal structure appears more stable than the tetragonal one.

Figures.

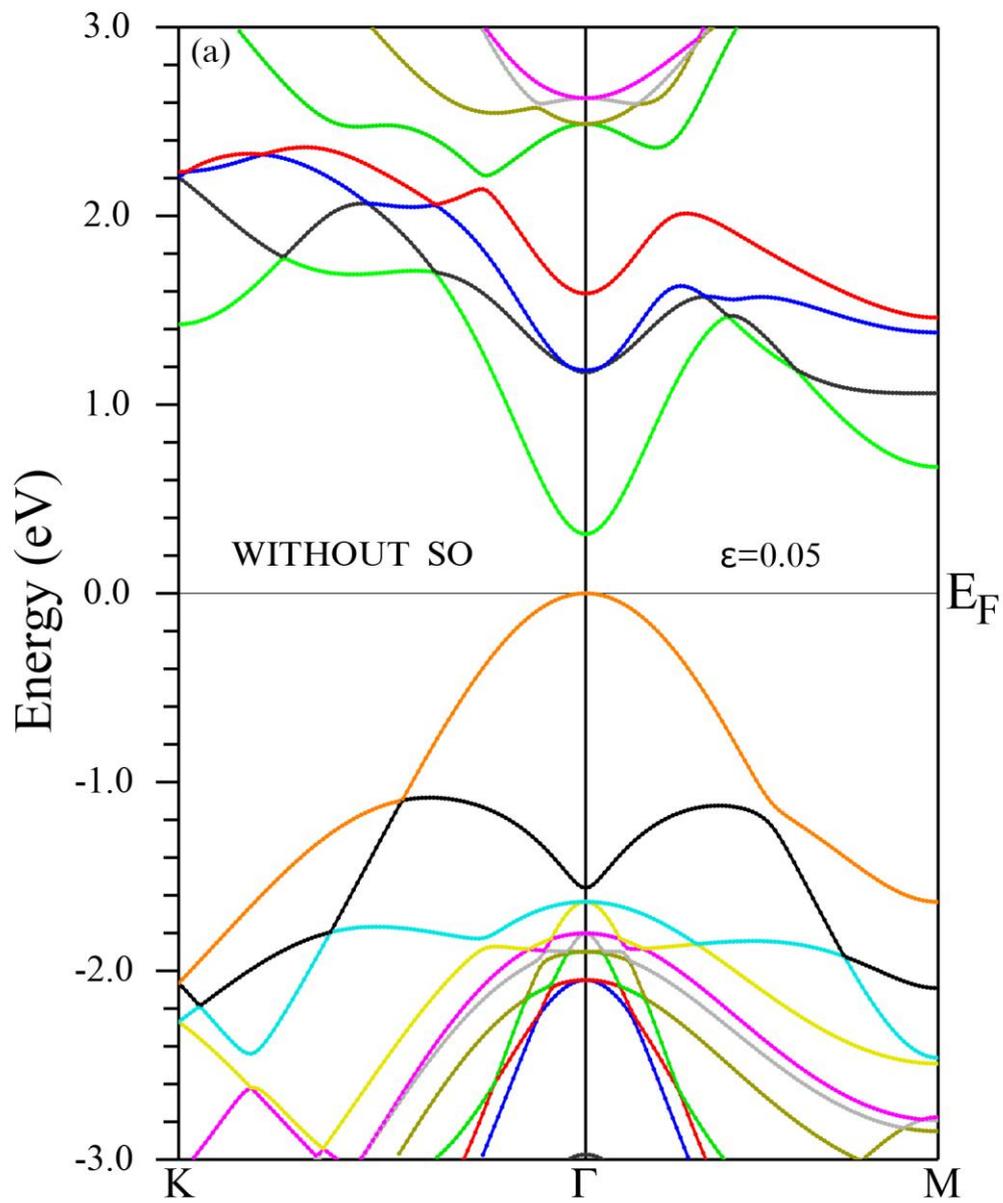

Fig. 1(a) Band structure of strained β-InTe with ε = 0.05, without SOC

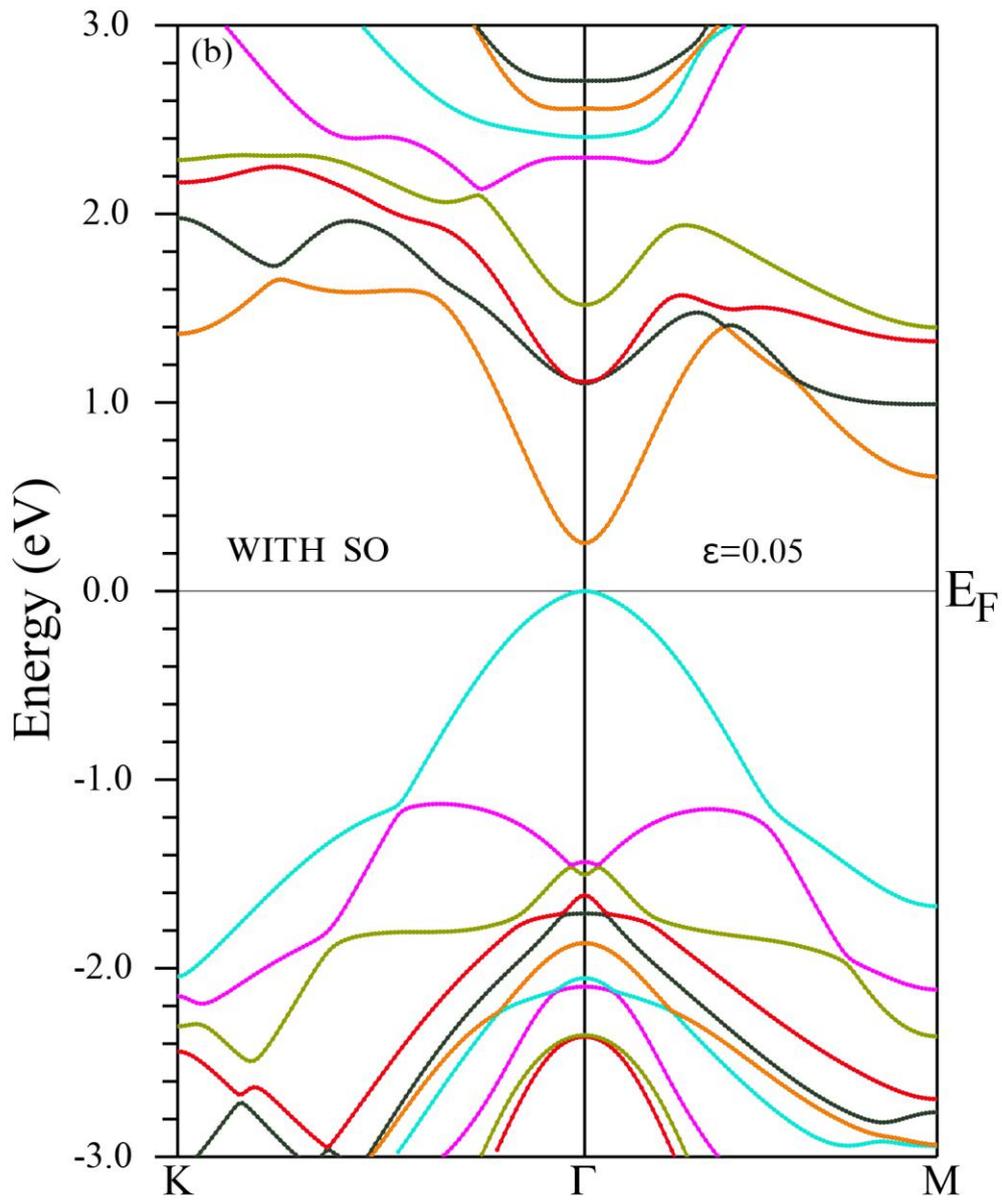

Fig 1(b) Band structure of strained β-InTe with ε = 0.05, with SOC

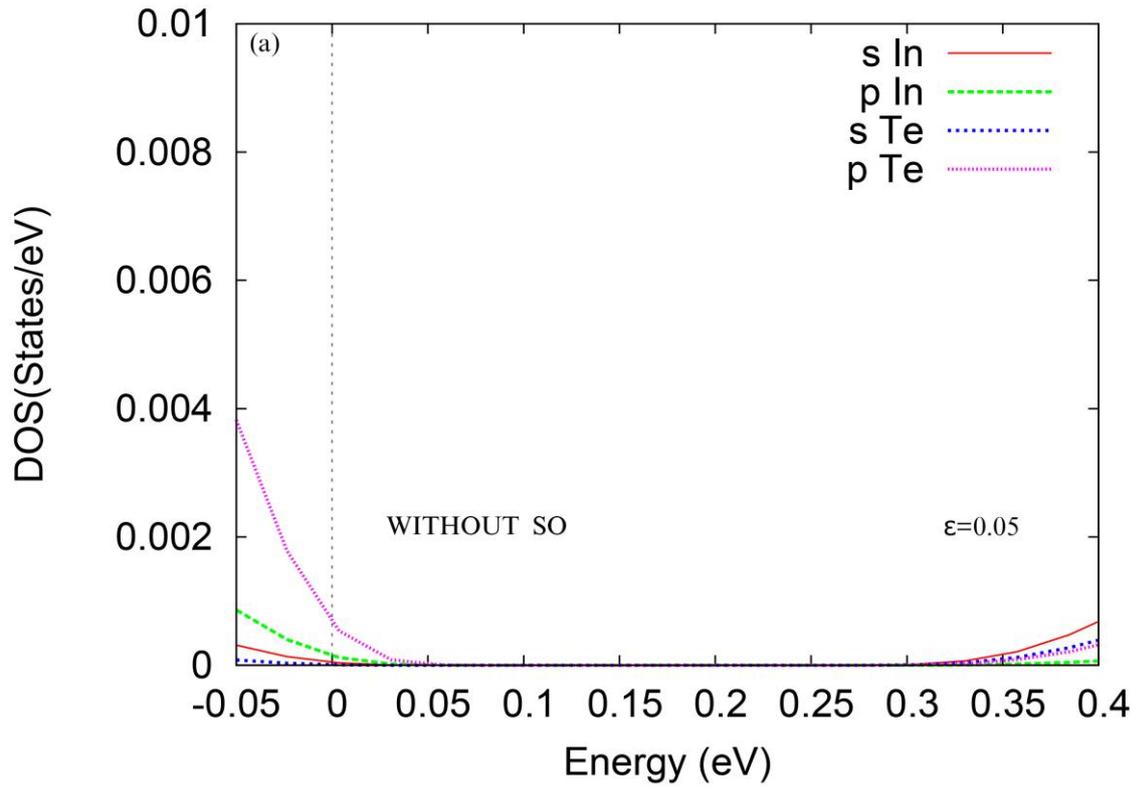

Fig. 2(a) Density of state of electrons in strained β-InTe with ε = 0.05, without SOC

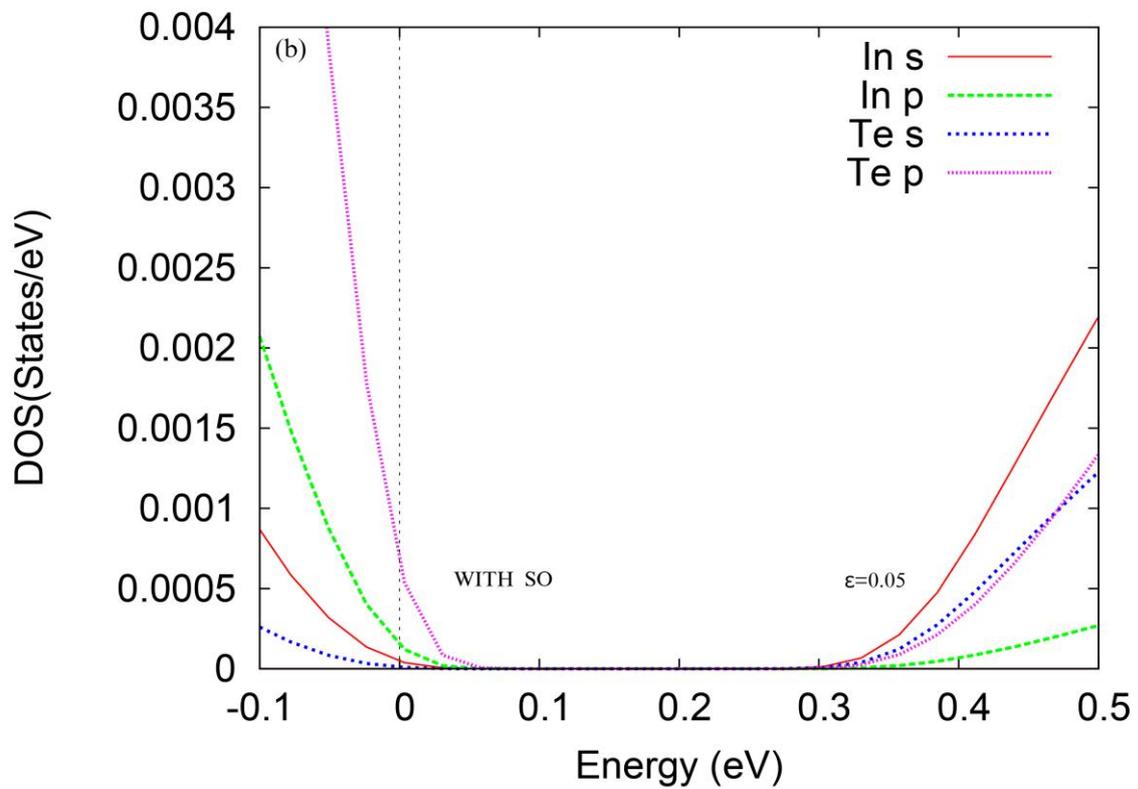

Fig. 2(b) Density of state of electrons in strained β-InTe with ε = 0.05, with SOC

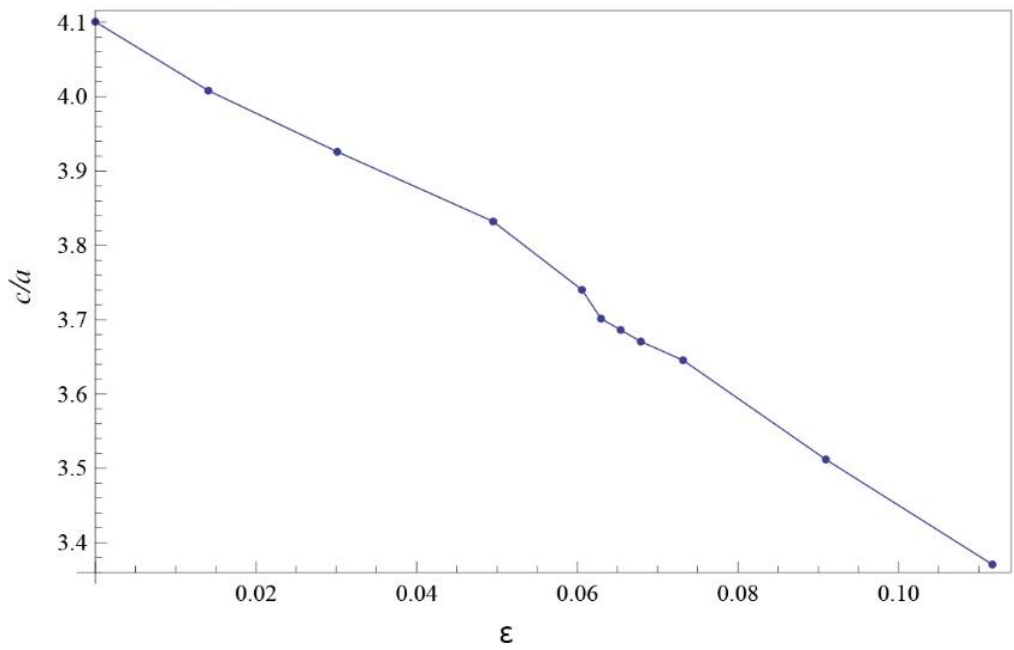

Fig. 3. $c/a$ versus strain magnitude ε.

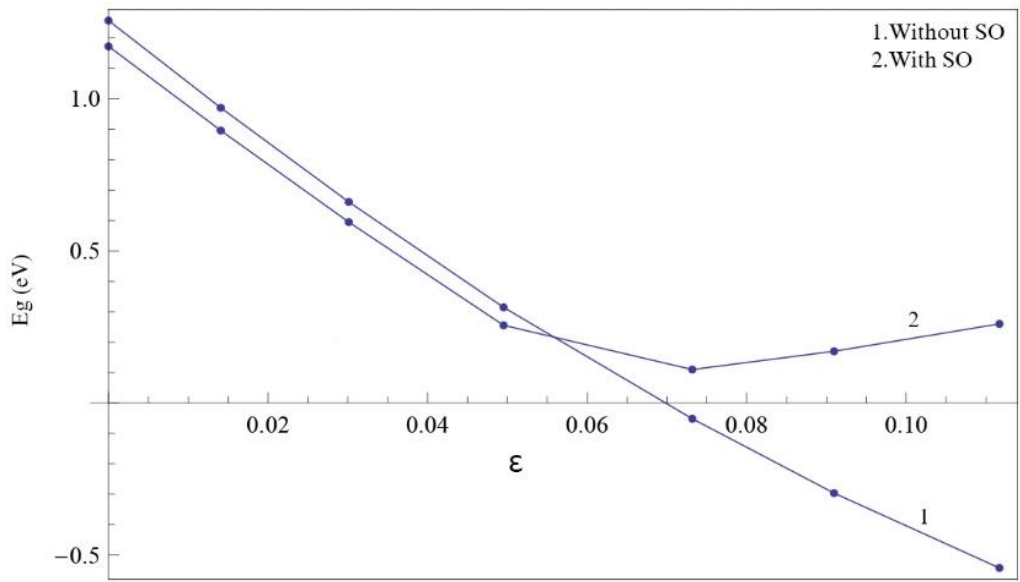

Fig. 4. $Eg$ versus strain magnitude ε.

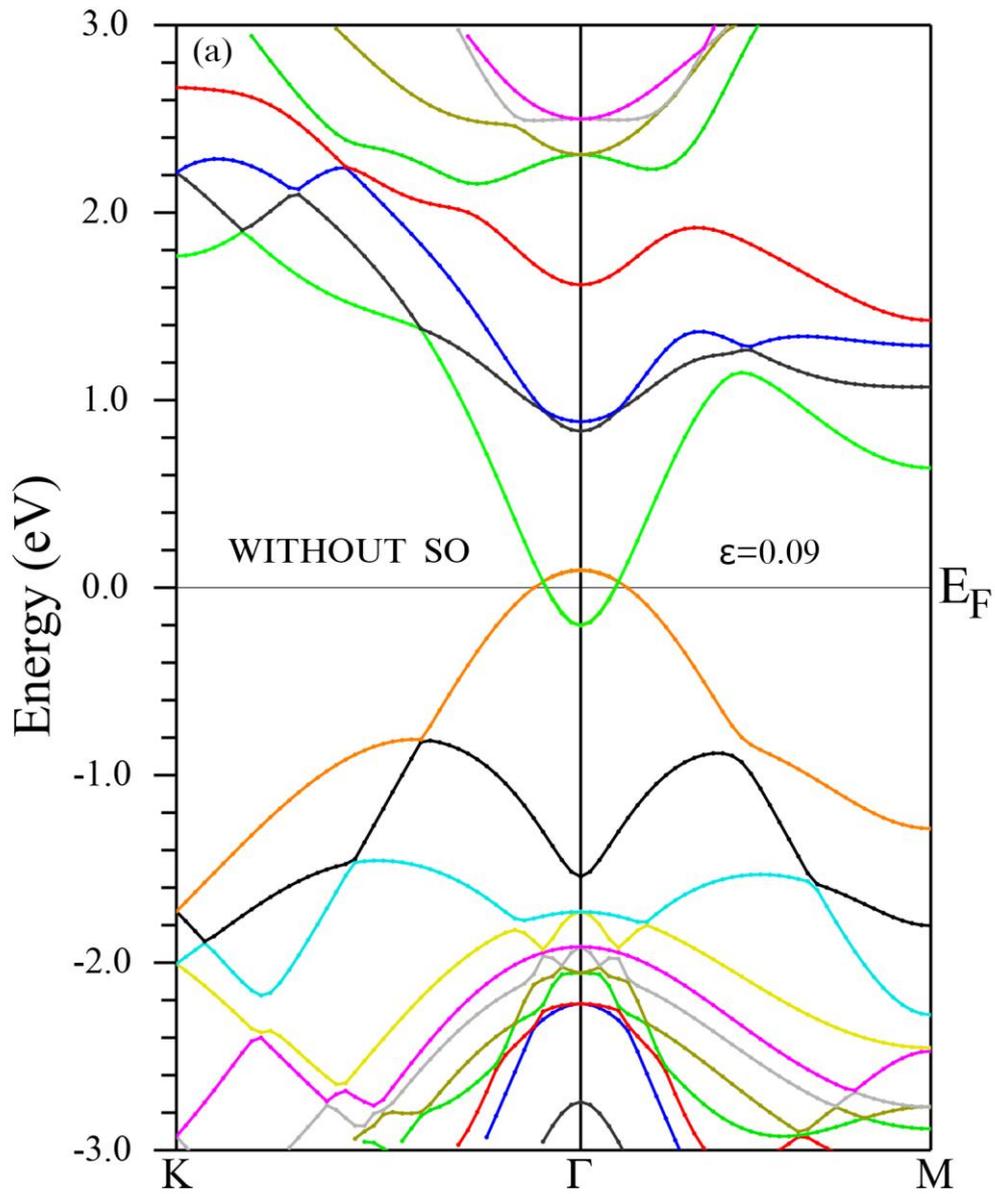

Fig. 5(a). Strained β-InTe band structure with ε = 0.09, without SOC.

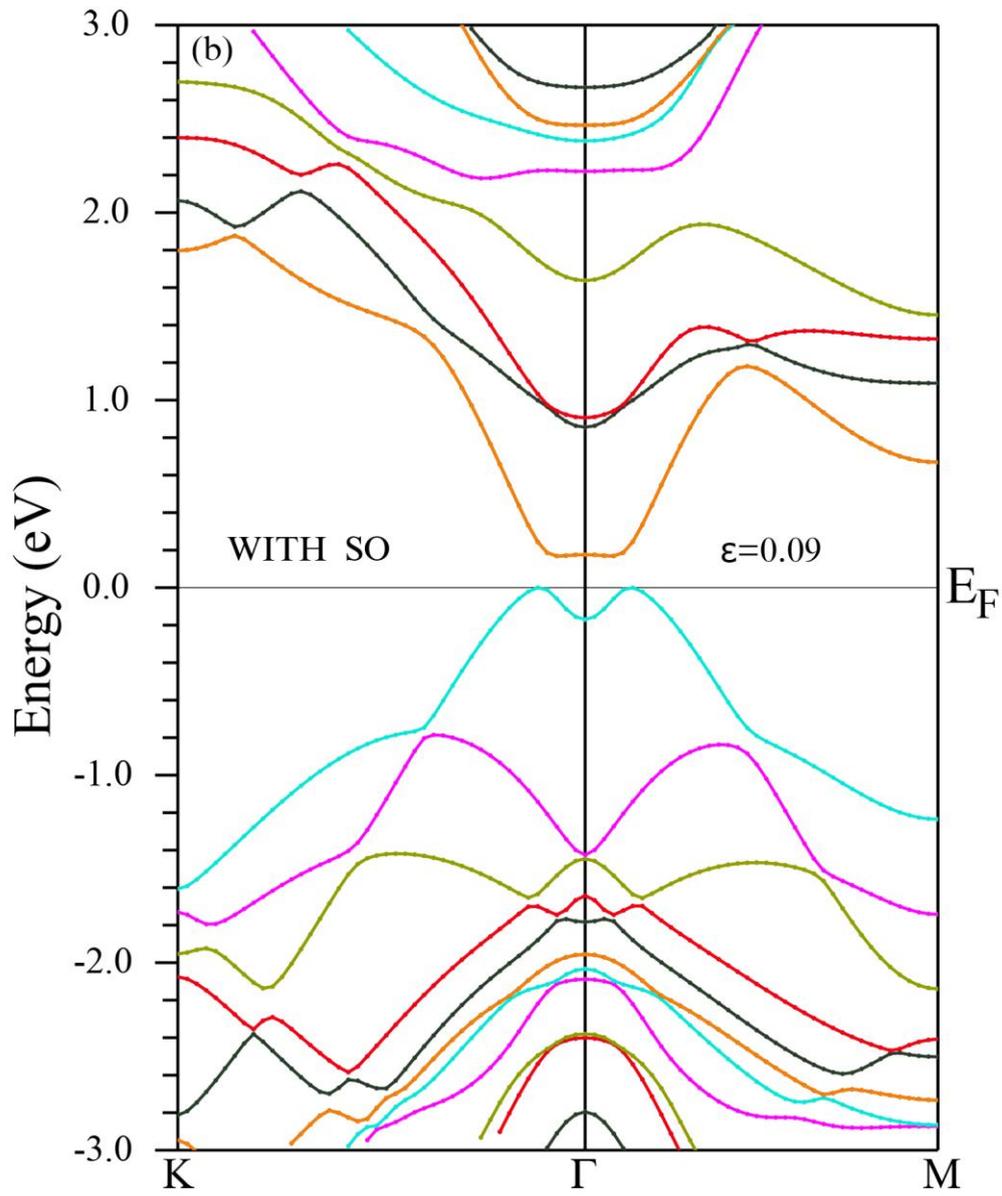

Fig. 5(b). Strained β-InTe band structure with $\varepsilon = 0.09$, with SOC.

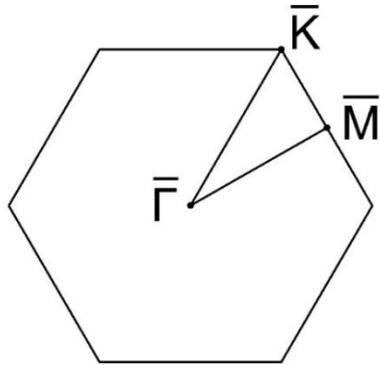

Fig. 6. Surface Brillouine zone of hexagonal crystal

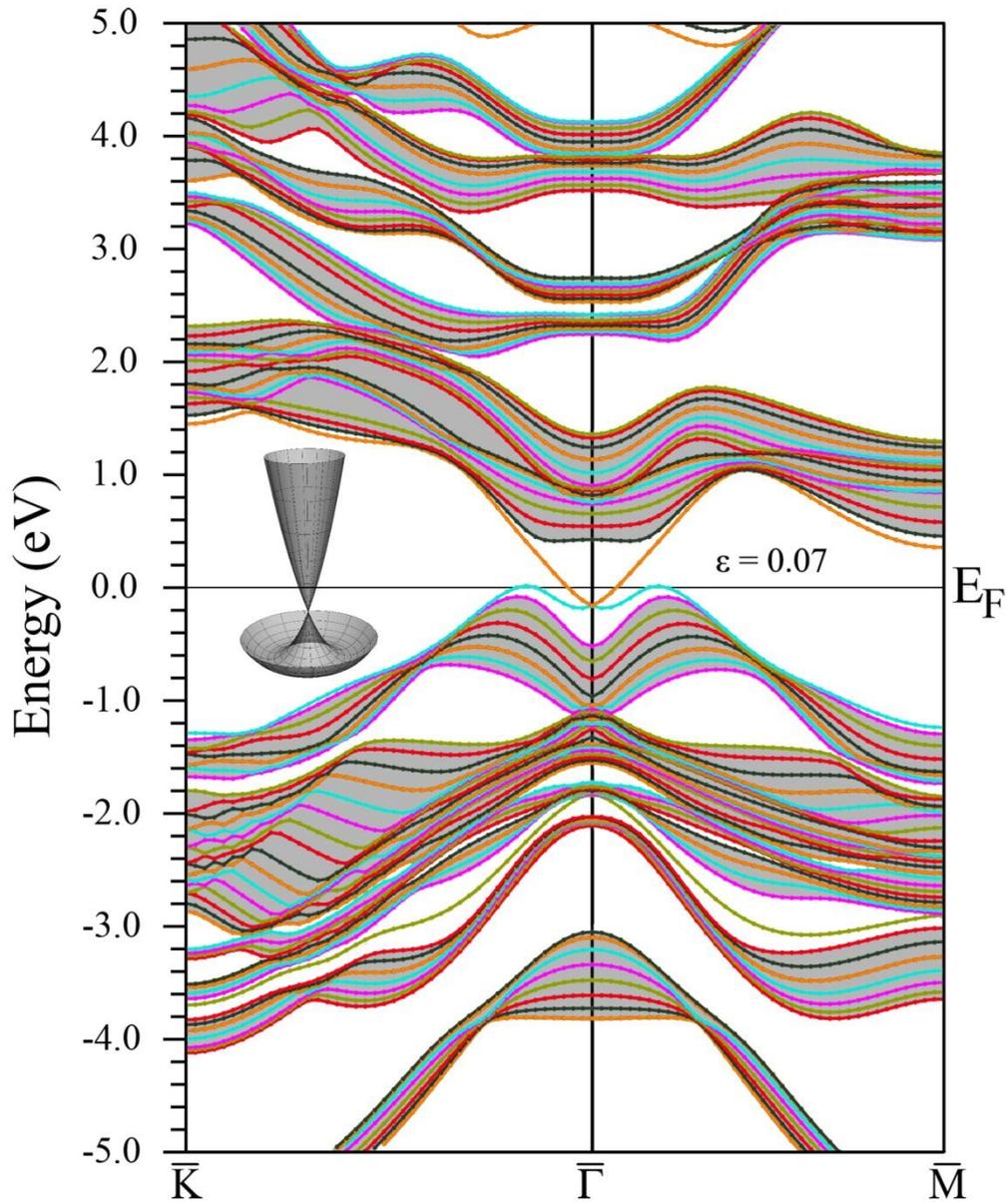

Fig. 7 Band structure of surface states and bulk states projected onto the surface Brillouine zone of strained β-InTe with $\varepsilon = 0.07$. The inset shows the Fermi surface near $\bar{\Gamma}$ (Dirac cone).

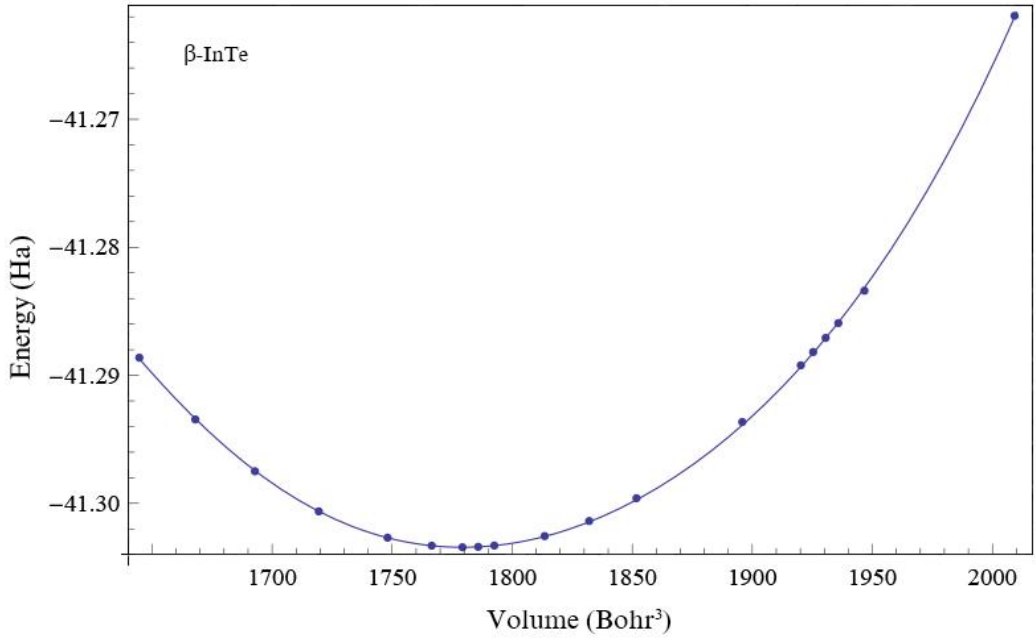

Fig. 8 Equation of state for β-InTe.

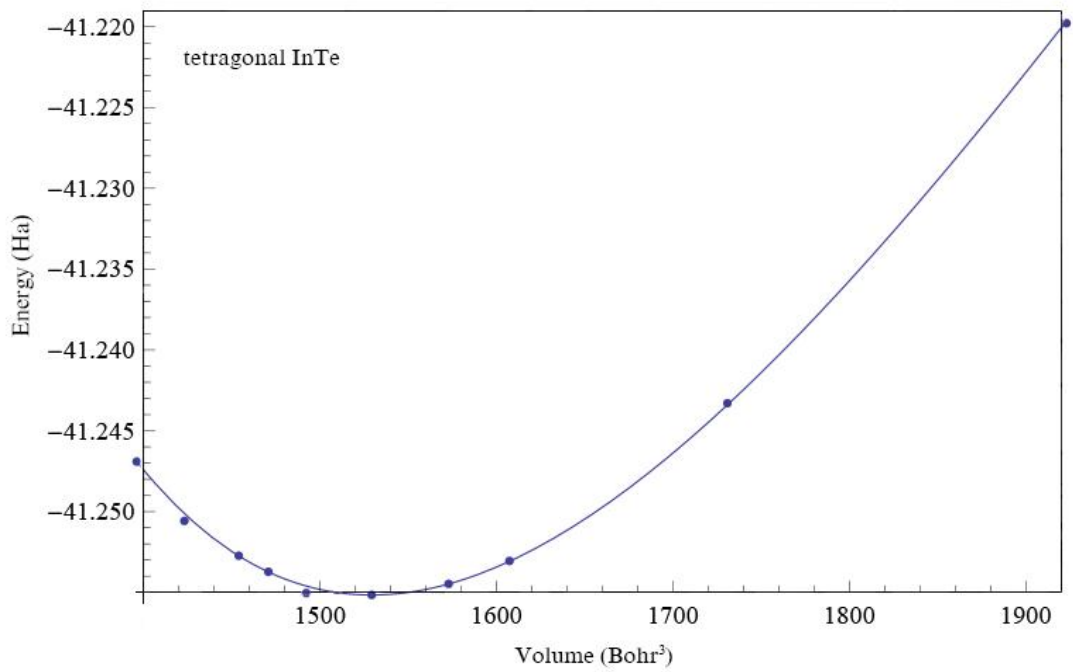

Fig. 9 Equation of state for tetragonal InTe.